\documentclass{aa}  

\usepackage{array}
\usepackage{graphicx}
\usepackage{amsmath}
\usepackage{txfonts}
\usepackage{soul}
\usepackage[normalem]{ulem}
\usepackage[colorlinks=true,linkcolor=blue, citecolor=blue]{hyperref}
\begin{document}

   \title{The Effect of Atmospheric Chemistry on the Optical Geometric Albedos of Hot Jupiters}

   \author{Kathryn D. Jones\inst{1} \and
          Brett M. Morris \inst{2} \and
          Kevin Heng \inst{3, 4, 5}
          }

\institute{
Center for Space and Habitability, University of Bern, Gesellschaftsstrasse 6, CH-3012 Bern, Switzerland \and
Space Telescope Science Institute, 3700 San Martin Dr, Baltimore, MD 21218, USA \and
Ludwig Maximilian University, University Observatory Munich, Scheinerstrasse 1, Munich D-81679, Germany \and
University of Warwick, Department of Physics, Astronomy \& Astrophysics Group, Coventry CV4 7AL, United Kingdom \and
University of Bern, ARTORG Center for Biomedical Engineering Research, Murtenstrasse 50, CH-3008, Bern, Switzerland 
}

   \authorrunning{K. D. Jones et al.}
 
   \date{Accepted 2026}
  \abstract
   {We investigate the geometric albedos of hot Jupiters by comparing observational data from space telescopes TESS, Kepler, CoRoT, and CHEOPS against theoretical models. The study aims to understand the distribution of observed geometric albedos across different bandpasses and how these observations align with or deviate from model predictions. We have curated a comprehensive sample of observed geometric albedos, using either existing Spitzer secondary eclipse measurements or a scaling law between the equilibrium and dayside temperature to remove any contaminating thermal planetary emission. We then utilised hierarchical Bayesian modelling to identify trends with planetary properties such as equilibrium temperature, gravity, and stellar metallicity. On a population level, we found no statistical difference in the distributions of geometric albedos measured by TESS compared to those by Kepler, CoRoT and CHEOPS. We confront the geometric albedo sample with a simple, but first principles, model that includes Rayleigh scattering by molecular hydrogen and absorption by sodium, water and titanium oxide and vanadium oxide. We find that the abundance of sodium and water are the key absorbers that influence the geometric albedos of hot Jupiters, whilst the addition of titanium oxide and vanadium oxide (in the absence of condensation) results in vanishing geometric albedos that are inconsistent with the observed distributions.} 


   \maketitle
%

\section{Introduction}

The geometric albedo of a planet, moon or exoplanet is the fraction of light reflected at full illumination with respect to the observer (at superior conjunction). Geometric albedo is a wavelength-dependent quantity and may be larger than unity if the incoming light is preferentially reflected back in the direction of the observer. In the Solar System, there is a long and rich history of measuring the geometric albedos of moons and planets using data measured from Voyager \citep[see, e.g.][]{Hanel1981} and Cassini \citep[see, e.g.][]{Pitman2010, Buratti2022}. For Jupiter, geometric albedo \textit{spectra} have been measured \citep[e.g.][]{Li2018}.

For exoplanets, the geometric albedo may be inferred from measurements of the eclipse depth ($D$), planetary radius ($R_p$) and semi-major axis ($a$) \citep{Seager2010},
\begin{equation}
    D = A_g \left( \frac{R_p}{a} \right)^2,
\end{equation}
where it is assumed that the light is being measured at a wavelength where thermal emission from the exoplanet is negligible and only reflected starlight is present \citep{Heng2013}.

In practice, the geometric albedos of exoplanets are measured over broad photometric bandpasses, rather than at specific wavelengths, by the CoRoT \citep[e.g.][]{Auvergne2009, Dang2018}, Kepler \citep[e.g.][]{Borucki2010, Morris2024, Heng2013, Heng2021b}, CHEOPS \citep[e.g.][]{Benz2021, Brandeker2022, Krenn2023} and TESS \citep[e.g.][]{Ricker2015, Wong2021} space telescopes. Almost all of these measurements are of hot Jupiters, due to their large eclipse depths and short periods leading to more significant eclipse depth measurements. Many hot Jupiters have been found to have low albedos, such as HD 189733b \citep[$0.076\pm0.016$,][]{Krenn2023} and HD 209458b \citep[$0.096\pm0.016$,][]{Brandeker2022}, while some show higher albedos, most likely due to reflective clouds, like Kepler-7 b \citep[$0.25^{+0.01}_{-0.02}$,][]{Demory2011a,Heng2021b,Morris2024}.

Atmospheric processes affect the reflectivity of exoplanet atmospheres. \cite{Sudarsky2000} predicted that hot Jupiters would have small geometric albedos because their high atmospheric temperatures may activate strong absorption of visible light by lines of sodium and potassium. \cite{Burrows2008} calculated albedo models for HD 209458b and HD 189733b and found a model without scattering clouds was able to reproduce the MOST observations \citep{Rowe2008}. \cite{Cahoy2010} modelled the albedos of Jupiter and Neptune analogues and found that clouds influenced the albedo spectra and temperature drove the effects of clouds more than metallicity. \cite{Madhusudhan2012} developed an analytic geometric albedo model framework including three different types of scattering phase functions to investigate the dependence of polarisation of Rayleigh scattering in the atmosphere on the planetary orbital inclination. In \cite{Mallonn2019}, they obtained geometric albedo limits for a number of ultra-hot Jupiters and found values consistent with low reflectivity in the optical to near-infrared. \cite{Heng2021b} derived a general analytic formula for the geometric albedo, which \cite{Brandeker2022} used to demonstrate that the $\sim 10\%$ geometric albedo of HD 209458b may be explained by scattering from molecular hydrogen and absorption from water and sodium.

The trends between different hot Jupiter properties and their geometric albedos have been a topic of study since the first occultation measurement of TrES-1b in 2005 \citep{Charbonneau2005}. Previous studies have included the set of hot Jupiters observed with Kepler \citep[see, e.g.][]{Heng2013, Angerhausen2015, Esteves2015}, and with TESS \citep{Wong2021}. 
They found a tentative positive correlation between dayside temperature and geometric albedo for the planets with $1500 < T_{\text{day}} < 3000$\,K. 

\begin{figure}
   \centering
   \includegraphics[width=\hsize]{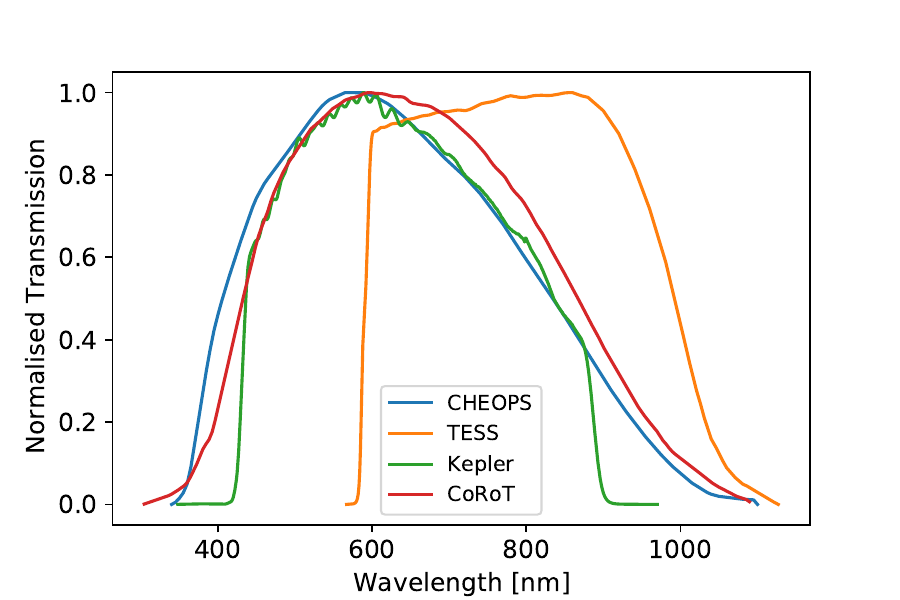}
      \caption{Transmission functions for CHEOPS, Kepler, CoRoT and TESS instruments as a function of wavelength. In this study we group CHEOPS, Kepler and CoRoT together due to their similarity in transmission functions. TESS is kept separate as it transmits more into the red.}
         \label{fig:trans_func}
\end{figure}

The growing, but heterogeneous, body of albedo measurements constrains the atmospheric physics of these planets. In this work, we investigate the distribution of observed geometric albedos from the TESS, CHEOPS, Kepler and CoRoT missions. We divide the four missions into two groups since CHEOPS, Kepler, and CoRoT (hereafter, CKC) have bandpasses with similar transmittance-weighted mean wavelengths ($\bar{\lambda} = 674, 664, 695$~nm, respectively), and the TESS bandpass has more transmittance in the red ($\bar{\lambda} = 797$~nm). The full transmission functions for the instruments can be seen in Figure \ref{fig:trans_func}.
We compare the observed albedos with the results from a simplified geometric albedo model \citep[derived from][]{Heng2021b} to answer the following questions:
\begin{itemize}
    \item Should we expect geometric albedo measurements in different filter bandpasses to have the same underlying distribution?

    \item Do geometric albedos correlate with equilibrium temperature, planetary surface gravity, stellar metallicity, or effective stellar temperature?

    \item Do correlations between albedo and planet properties dependent on the photometric bandpass of the measurements (e.g. Kepler, TESS)?

    \item Does the scatter in albedo observations set constraints on the atmospheric properties of hot Jupiters?

\end{itemize}

In Section \ref{sec:thermaldecom} we describe our data curation methods and the procedure we followed to thermally decontaminate the albedos of targets where that had not been done previously. The curated albedo measurements are shown in Table \ref{tab:albedos}. Section \ref{sec:modellingAg} details our theoretical geometric albedo model and the prior assumptions used to produce 
the albedo distributions we investigate in Section \ref{sec:model}. In Section \ref{sec:obs} we investigate whether the observational data is correlated with other system or atmospheric parameters, and to what significance. We also look at the variations between the TESS and CKC datasets. In Section \ref{sec:flowchart} we synthesise our findings from both the observations and the modelling and present a flowchart to show what information can be determined from measuring the geometric albedo of a planet. 

\section{Methods}

\subsection{Data curation and thermal decontamination}
\label{sec:thermaldecom}

To determine if the measured geometric albedos display any trends with the stellar or exoplanetary properties, we first need to curate and collect them. We divide geometric albedo measurements into two categories: those from the CoRoT, Kepler or CHEOPS missions, which have similar optical filters; and the TESS mission, with measurements at slightly longer optical wavelengths.  

The observed flux coming from the planet can be a combination of reflected starlight and thermal emission from the planet itself. The process of removing thermal emission to compute the albedo is called thermal decontamination \citep[see, e.g. ][]{Heng2013}. Thermal decontamination was already performed by \cite{Wong2021} for the TESS observations. There were a few objects where only an upper limit was reported. For these planets, we reapplied the thermal decontamination (with the same method as in \citealt{Wong2021}), given the eclipse depth and dayside temperature and kept the full error bars, even if they went below zero, to reduce the statistical bias towards $A_g > 0$. We exclude KELT-1b from consideration, as it is a brown dwarf. We also exclude Kepler-13Ab, as \cite{Wong2021} have shown that the inferred TESS geometric albedo depends heavily on the assumed infrared absorbers.

Many of the CoRoT and Kepler albedos in the literature have not been thermally decontaminated. For these albedos, we decontaminate with:
\begin{equation}
    A_g = D \left( \frac{a}{R_p} \right)^2 - \frac{\int F_\lambda \mathcal{T} \, d\lambda}{\int F_{\star,\lambda} \mathcal{T} \, d\lambda} \left( \frac{a}{R_\star} \right)^2,
\end{equation}
where $F_\lambda$ is the planetary thermal dayside flux as a function of wavelength, $F_{\star,\lambda}$ is the stellar flux as a function of wavelength, and $\mathcal{T}$ is the telescope transmission function. Since these albedos are measured in wide bandpasses, we assume a blackbody for the thermal emission of the planet ($F_\lambda = \pi B_\lambda(T_{\text{day}}$)), with an integrated dayside hemisphere temperature. Where available, we use dayside temperatures derived from previous Spitzer observations. Many of the CoRoT and Kepler objects have no Spitzer measurements, so we estimate the dayside temperatures with the scaling relationship between $T_{\rm day}$ and the equilibrium temperature ($T_{\rm eq}$) calibrated by \cite{Beatty2019}. We account only for uncertainties on the measured eclipse depths and ignore the uncertainties on $T_\star$, $a/R_\star$ and $T_{\rm day}$, because the joint posterior distributions of these quantities are generally unavailable in the literature. The list of decontaminated geometric albedo values are reported in Table \ref{tab:albedos}.

\subsection{Modelling the bandpass-integrated geometric albedo}
\label{sec:modellingAg}

The geometric albedo may be calculated from first principles, if the single-scattering albedo ($\omega$) is known, using the analytical formula derived by \cite{Heng2021b},
\begin{equation}
    A_g = \frac{\omega}{8}(P_0 - 1) + \frac{\epsilon}{2} + \frac{\epsilon^2}{6} + \frac{\epsilon^3}{24}, 
\end{equation}
where $\epsilon=(1-\gamma)/(1+\gamma)$ and $\gamma=\sqrt{1-\omega}$. $P_0$ is a constant which describes the amount of single-scattering. This formula was derived for a semi-infinite atmosphere \citep{Chandrasekhar1960}, where the optical depth (but not the spatial distance) transitions from zero to infinity. We construct the single-scattering albedo $\omega$ with
\begin{equation}
    \omega = \frac{0.9\sigma_{H_2}}{0.9\sigma_{H_2} + \sum X_{M}\sigma_{M}},
\label{eqn:singlescatalb}
\end{equation}

where $X_{M}$ is the volume mixing ratio (relative abundance by number) of a chemical species $M$ (e.g. water, sodium), and $\sigma_{M}$ is the opacity of $M$. We assume a hydrogen-dominated atmosphere consistent with the cosmic abundance of hydrogen and helium, but ignore the scattering and absorption by helium. We assume that the other species have a total mixing ratio well below $1\%$. We compute mixing ratios $X_M$ with \texttt{FastChem} \citep{Stock2018,Stock2022,Kitzmann2024} for several hundred species in chemical equilibrium given the temperature, pressure and metallicity. We retrieve atomic and molecular opacities $\sigma_M$ from \texttt{DACE}\footnote{\url{https://dace.unige.ch}}, which were computed with HELIOS-K \citep{Grimm2015, Grimm2021}. We access opacities and mixing ratios from \texttt{DACE} and \texttt{FastChem} using the radiative transfer package \texttt{shone} (Morris et al., in prep.)\footnote{\url{https://github.com/bmorris3/shone}}.

To compute the bandpass-integrated geometric albedo, one needs to integrate the geometric albedo over the bandpass filter, weighted by the stellar spectral flux,
\begin{equation}
    \bar{A_g} = \frac{\int A_g(\lambda) F_{\star,\lambda} \mathcal{T}\,d\lambda}{\int F_{\star,\lambda} \mathcal{T} \,d\lambda}.
\end{equation}
We note here that this equation is correct using the TESS transmission filter due to it being an energy counter, however for CoRoT, Kepler and CHEOPS, an additional factor of $\lambda$ must be applied to both the numerator and denominator as they are photon counters \citep{Rodrigo2020}. In our model, for a given stellar effective temperature, we use a PHOENIX spectrum to model to stellar spectral flux $F_{\star, \lambda}$ \citep{Husser2013}. 

We begin with a range of metallicities, temperatures and pressures given the measured properties of the planets in Table \ref{tab:albedos}, and assume chemical equilibrium to calculate the volume mixing ratios of each species. Planets at high temperatures are more likely to be near chemical equilibrium, but the threshold effective temperature at which this approximation breaks down is not easily specified. We parameterise the gas-phase chemistry by a metallicity, which assumes that the ratios of elemental abundances are locked to their solar values. In practice, chemical equilibrium results ratios of water to sodium that depend only on the assumed temperature and pressure. We also run a second model with water and sodium abundances that are allowed to deviate from chemical equilibrium.

In this investigation we include species that are: (a) expected to be present in hot Jupiters within the range of equilibrium temperatures in Table~\ref{tab:albedos}; and (b) have significant absorption bands in the optical. We include $\textrm{H}_2\textrm{O}$ and Na, along with Rayleigh scattering from $\textrm{H}_2$, and test the effects of TiO and VO.

The range of stellar effective temperatures $T_\star$ is guided by the minimum and maximum values corresponding to the objects of our sample: 4550 to 8000\,K. We need both the stellar temperature, and also the temperature of the planet at optical depth $\tau\sim1$ in the atmosphere. The relationship between these temperatures is complex and not known in general, so we sample these two temperatures independently. We use a range of values of the temperatures equal to the range of equilibrium temperatures in our sample (between 1300 and 2700\,K), where equilibrium temperature is defined as
\begin{equation}
    T_{\rm eq} = T_\star \sqrt{\frac{R_*}{2a}}, 
    \label{eqn:teq}
\end{equation}
where $R_*$ is the stellar radius and and $a$ is the semi-major axis. Equilibrium temperature is not the same as the dayside temperature of the planet, however in our simple model, it is enough to guide our chosen temperature range.

We set a prior on metallicity $\log\mathcal{U}(-1,1)$ based off an empirical scaling relation between metallicity and planetary mass \citep{Swain2024, Wakeford2017}. The range of pressures used follows the estimated photospheric pressures of hot Jupiters.

\section{Geometric albedo observations}
\label{sec:obs}

\subsection{Observed geometric albedo trends}
The set of observed geometric albedos are reported in Table \ref{tab:albedos}. In total there are 17 targets in the CKC band and 19 in the TESS band. We divide the objects into two section: those in the top section have their albedos taken directly from previous analyses (where thermal decontamination has been carried out); and in the bottom half, we used eclipse depths and dayside temperatures from the enumerated references to decontaminate the geometric albedos (see method in Section \ref{sec:thermaldecom}). All of the targets are within the hot to ultra-hot Jupiter regime with equilibrium temperatures ranging from $1179\pm 11$\,K to $2764\pm 37$\,K. 

\begin{table*}
\caption{Collection of hot Jupiter exoplanets with their reported geometric albedo ($A_g$) and other stellar and planetary parameters.}
\begin{tabular}{cccccccc}
\hline
\hline
Object      & $T_*$ [K] & Fe/H  & $T_{\text{eq}}$ [K]& log(g) [cgs]  & a/$R_*$  & Band   & $A_g $ \\
 \hline
CoRotT-2b   & $5625\pm120$\tablefootmark{a}   & $0.0\pm0.1$\tablefootmark{a}    & $1556\pm27$  & $3.58\pm0.03$\tablefootmark{e}  & $6.70\pm0.03$\tablefootmark{e}   &  Corot  & $0.08^{+0.08}_{-0.04}$\tablefootmark{f}   \\
HAT-P-7b     & $6350\pm80$\tablefootmark{a}    & $0.26\pm0.08$\tablefootmark{a}  & $2264\pm21$  & $3.30\pm0.09$\tablefootmark{g}  & $4.1545^{+0.0029}_{-0.0025}$\tablefootmark{g}  & Kepler & $0.09\pm0.02$\tablefootmark{h}\\
Kepler-7b   & $5933\pm50$\tablefootmark{a}    & $0.11\pm0.05$\tablefootmark{a}  & $1670\pm17$  & $2.612\pm0.052$\tablefootmark{a}  & $6.637\pm0.021$\tablefootmark{g}    &  Kepler & $0.25^{+0.01}_{-0.02}$\tablefootmark{m}  \\
Kepler-41b  & $5750\pm100$\tablefootmark{a}   & $0.38\pm0.11$\tablefootmark{a}  & $1808\pm32$  & $2.926\pm0.057$\tablefootmark{a}  & $5.053\pm0.021$\tablefootmark{g}  & Kepler & $0.13^{+0.01}_{-0.02}$\tablefootmark{h}   \\
Qatar-1b    & $5013^{+93}_{-88}$\tablefootmark{k}  & $0.171\pm0.096$\tablefootmark{k}   & $1400\pm28$  & $3.39\pm0.03$\tablefootmark{k}  & $6.41^{+0.11}_{-0.10}$\tablefootmark{i}    &  TESS   & $0.14 \pm 0.11 $ \tablefootmark{i}   \\
TrES-2b     & $5850\pm50$\tablefootmark{a}    & $-0.15\pm0.1$\tablefootmark{a}  & $1458\pm18$  & $3.30\pm0.03$\tablefootmark{a}  & $7.903^{+0.019}_{-0.016}$\tablefootmark{g}  &  Kepler & $0.01^{+0.00}_{-0.01}$\tablefootmark{h}   \\
TrES-3b     & $5650\pm75$\tablefootmark{a}    & $-0.19\pm0.08$\tablefootmark{a} & $1613\pm50$  & $3.40\pm0.03$\tablefootmark{a}  & $5.82^{+0.12}_{-0.13}$\tablefootmark{j}   & TESS   & $0.14 \pm 0.13$\tablefootmark{i}    \\
WASP-4b     & $5436\pm34$\tablefootmark{a}    & $-0.05\pm0.04$\tablefootmark{a}  & $1648\pm13$  & $3.213\pm0.098$\tablefootmark{a}  & $5.438^{+0.044}_{-0.057}$\tablefootmark{q}  & TESS   & $0.09 \pm 0.09$\tablefootmark{q}    \\
WASP-5b     & $5770\pm65$\tablefootmark{a}    & $0.09\pm0.04$\tablefootmark{a}  & $1762\pm41$  & $3.455\pm0.043$\tablefootmark{a}  & $5.36\pm0.22$\tablefootmark{q}     &  TESS   & $0.0002^{+0.166}_{-0.174}$\tablefootmark{q}   \\ 
WASP-12b    & $6250\pm100$\tablefootmark{a}   & $0.32\pm0.12$\tablefootmark{a}  & $2526\pm48$  & $3.015\pm0.059$\tablefootmark{a}  & $3.062^{+0.063}_{-0.066}$\tablefootmark{i}   & TESS   & $0.13 \pm0.06$\tablefootmark{i}  \\
WASP-19b    & $5500\pm100$\tablefootmark{a}   & $0.02\pm0.09$\tablefootmark{a}  & $2055\pm42$  & $3.222\pm0.048$\tablefootmark{a}  & $3.582^{+0.074}_{-0.067}$\tablefootmark{q}  & TESS   & $0.17 \pm 0.07$\tablefootmark{q}    \\
WASP-33b    & $7430\pm100$\tablefootmark{r}   & $0.1\pm0.2$\tablefootmark{r}    & $2764\pm37$  & $3.459\pm0.098$\tablefootmark{t}  & $3.614\pm0.009$\tablefootmark{s}  & TESS   & $-0.04\pm 0.04 $\tablefootmark{s}  \\
WASP-36b    & $5900\pm150$\tablefootmark{a}   & $-0.26\pm0.1$\tablefootmark{a}  & $1738\pm59$  & $3.540\pm0.023$\tablefootmark{a}  & $5.76^{+0.26}_{-0.27}$\tablefootmark{q}  & TESS   & $0.16\pm 0.15$\tablefootmark{q}  \\
WASP-43b    & $4798\pm216$\tablefootmark{u}   & $-0.13\pm0.08$\tablefootmark{u} & $1559\pm71$  & $3.675\pm0.019$\tablefootmark{a}  & $4.734^{+0.054}_{-0.053}$\tablefootmark{q}  & TESS   & $0.13\pm 0.06 $\tablefootmark{q} \\
WASP-46b    & $5725\pm39$\tablefootmark{u}    & $-0.18\pm0.03$\tablefootmark{u} & $1630\pm36$  & $3.533\pm0.036$\tablefootmark{a}  & $6.17^{+0.28}_{-0.24}$\tablefootmark{q}  & TESS   & $0.38\pm0.27$\tablefootmark{q}    \\
WASP-64b    & $5550\pm150$\tablefootmark{a}   & $-0.08\pm0.11$\tablefootmark{a} & $1669\pm54$  & $3.272\pm0.038$\tablefootmark{a}  & $5.53^{+0.14}_{-0.25}$\tablefootmark{q}  & TESS   & $0.38\pm 0.26$\tablefootmark{q}    \\
WASP-77Ab   & $5500\pm80$\tablefootmark{a}    & $0.00\pm0.11$\tablefootmark{a}  & $1712\pm30$  & $3.475\pm0.020$\tablefootmark{a}  & $5.162^{+0.120}_{-0.080}$\tablefootmark{q}  & TESS   & $0.06\pm 0.05$\tablefootmark{q}  \\
WASP-100b   & $6900\pm120$\tablefootmark{a}   & $-0.03\pm0.1$\tablefootmark{a}  & $2102\pm39$  & $3.24\pm0.14$\tablefootmark{a}    & $5.389\pm0.064$\tablefootmark{q}   & TESS   & $0.22\pm0.08$\tablefootmark{q}    \\
WASP-121b   & $6460\pm140$\tablefootmark{v}   & $0.13\pm0.09$\tablefootmark{v}   & $2339\pm51$  & $2.970\pm0.017$\tablefootmark{w}  & $3.815^{+0.018}_{-0.032}$\tablefootmark{q}  & TESS   & $0.26\pm 0.06$ \tablefootmark{q}   \\
HD209458b   & $6065\pm50$\tablefootmark{a}    & $0.00\pm0.05$\tablefootmark{a}  & $1445\pm13$  & $2.958\pm0.013$\tablefootmark{a}  & $8.807\pm0.051$\tablefootmark{z}   & CHEOPS & $0.096\pm 0.016$\tablefootmark{z}   \\
HD189733b   & $4969\pm48$\tablefootmark{$\alpha$}  & $-0.08\pm0.03$\tablefootmark{$\alpha$}  & $1179\pm11$ & $3.332\pm0.026$\tablefootmark{a}  & $8.8843\pm0.0175$\tablefootmark{$\beta$}  & CHEOPS & $0.076\pm 0.016$\tablefootmark{$\beta$} \\ 
\hline
CoRoT-1b    & $6298 \pm 66$\tablefootmark{a}  & $0.06\pm0.07$\tablefootmark{a}  & $2039\pm24$  & $3.06\pm0.07$\tablefootmark{b} & $4.92\pm0.08$\tablefootmark{b} & Corot  & $0.064^{+0.110}_{-0.139}$\tablefootmark{c,d}  \\
HAT-P-7b     & $6350\pm80$\tablefootmark{a}    & $0.26\pm0.08$\tablefootmark{a}  & $2264\pm21$  & $3.30\pm0.09$\tablefootmark{g}  & $4.1545^{+0.0029}_{-0.0025}$\tablefootmark{g}   & TESS   & $0.078^{+0.102}_{-0.103}$\tablefootmark{i}   \\
Kepler-5b   & $6297\pm60$\tablefootmark{a}    & $0.04\pm0.06$\tablefootmark{a}  & $1763\pm17$  & $3.410\pm0.034$\tablefootmark{a}  & $6.450^{+0.021}_{-0.025}$\tablefootmark{g}   & Kepler & $0.097 \pm0.037$\tablefootmark{g,l}   \\
Kepler-6b   & $5647\pm44$\tablefootmark{a}    & $0.34\pm0.04$\tablefootmark{a}  & $1461\pm16$  & $3.062\pm0.018$\tablefootmark{a}  & $7.503\pm0.022$\tablefootmark{g}    &  Kepler & $0.059 \pm0.035$\tablefootmark{g,l}   \\
Kepler-8b   & $6210\pm150$\tablefootmark{a}   & $-0.055\pm0.03$\tablefootmark{a}  & $1669\pm16$  & $2.84\pm0.12$\tablefootmark{a}  & $6.854^{+0.018}_{-0.017}$\tablefootmark{g}  & Kepler & $0.109^{+0.050}_{-0.052}$\tablefootmark{g,n}  \\
Kepler-12b  & $5950\pm100$\tablefootmark{a}   & $0.07\pm0.04$\tablefootmark{a}  & $1480\pm15$  & $2.572\pm0.045$\tablefootmark{a}  & $8.019^{+0.014}_{-0.013}$\tablefootmark{g}  & Kepler & $0.084\pm 0.040$\tablefootmark{g,o}    \\
Kepler-43b  & $6050\pm100$\tablefootmark{a}   & $0.4\pm0.1$\tablefootmark{a}    & $1620\pm27$  & $3.766\pm0.028$\tablefootmark{a}  & $6.975^{+0.041}_{-0.047}$\tablefootmark{g} & Kepler & $0.036\pm 0.250$\tablefootmark{g,n}    \\
Kepler-76b  & $6409\pm95$\tablefootmark{a}    & $-0.1\pm0.2$\tablefootmark{a}   & $2145\pm33$  & $3.499\pm0.082$\tablefootmark{a}  & $4.464^{+0.041}_{-0.049}$\tablefootmark{g}   & Kepler & $0.148^{+0.028}_{-0.025}$\tablefootmark{g,n}   \\
Kepler-91b  & $4550\pm75$\tablefootmark{p}    & $0.11\pm0.07$\tablefootmark{p}   & $2036\pm38$  & $3.0\pm0.1$\tablefootmark{g}    & $2.496^{+0.043}_{-0.050}$\tablefootmark{g}  & Kepler & $0.362^{+0.250}_{-0.234}$\tablefootmark{g,n}   \\
Kepler-412b & $5750\pm90$\tablefootmark{a}    & $0.27\pm0.12$\tablefootmark{a}  & $1848\pm29$  & $3.117\pm0.043$\tablefootmark{a}  & $4.841^{+0.023}_{-0.024}$\tablefootmark{g}  & Kepler & $0.066^{+0.047}_{-0.042}$\tablefootmark{g,n}   \\
WASP-3b     & $6400\pm100$\tablefootmark{a}   & $0.0\pm0.2$\tablefootmark{a}    & $1947\pm46$  & $3.368\pm0.039$\tablefootmark{a}  & $5.40\pm0.19$\tablefootmark{i} & TESS   & $0.181^{+0.190}_{-0.197}$\tablefootmark{i}  \\
WASP-5b     & $5770\pm65$\tablefootmark{a}    & $0.09\pm0.04$\tablefootmark{a}  & $1762\pm41$  & $3.455\pm0.043$\tablefootmark{a}  & $5.36\pm0.22$\tablefootmark{q}     &  TESS   & $0.0002^{+0.166}_{-0.174}$\tablefootmark{q}   \\ 
WASP-18b    & $6400\pm100$\tablefootmark{a}   & $0.00\pm0.09$\tablefootmark{a}  & $2406\pm40$  & $4.323\pm0.058$\tablefootmark{a}  & $3.539^{+0.039}_{-0.035}$\tablefootmark{q}  & TESS   & $-0.038^{+0.076}_{-0.070}$\tablefootmark{q}  \\
WASP-78b    & $6100\pm150$\tablefootmark{a}   & $-0.35\pm0.14$\tablefootmark{a} & $2219\pm59$  & $2.903\pm0.064$\tablefootmark{a}  & $3.778^{+0.060}_{-0.098}$\tablefootmark{q}  & TESS & $0.189^{+0.207}_{-0.211}$\tablefootmark{q}   \\
WASP-189b   & $8000\pm80$\tablefootmark{x}    & $0.29\pm0.13$\tablefootmark{x}   & $2641\pm28$  & $3.27\pm0.05$\tablefootmark{x}    & $4.587^{+0.037}_{-0.034}$\tablefootmark{y}   & CHEOPS & $0.202^{+0.046}_{-0.048}$\tablefootmark{n,y}  

\end{tabular}
\tablefoot{Objects in the top half of the table have their albedos taken directly from previous analyses (where thermal decontamination has been carried out). For objects in the lower half, using measured eclipse depths and dayside temperatures from the references specified, we have calculated the thermally decontaminated geometric albedos (see method in Section \ref{sec:thermaldecom}). The 'Band' column details which bandpass the planet's eclipse depth was measured in, and therefore which bandpass the geometric albedo corresponds to. $T_{\text{eq}}$ refers to the equilibrium temperature of the planet, calculated from the other values in this table using Equation \ref{eqn:teq}. 
\tablefoottext{a}{\cite{Bonomo2017}}, \tablefoottext{b}{\cite{Barge2008}}, \tablefoottext{c}{\cite{Deming2011}}, \tablefoottext{d}{\cite{Alonso2009}}, \tablefoottext{e}{\cite{Alonso2008}}, \tablefoottext{f}{\cite{Dang2018}}, \tablefoottext{g}{\cite{Esteves2015}}, \tablefoottext{h}{\cite{Morris2024}}, \tablefoottext{i}{\cite{Wong2021}}, \tablefoottext{j}{\cite{Patel2022}}, \tablefoottext{k}{\cite{Collins2017}}, \tablefoottext{l}{\cite{Desert2011}}, \tablefoottext{m}{\cite{Heng2021b}}, \tablefoottext{n}{\cite{Beatty2019}}, \tablefoottext{o}{\cite{Fortney2011}}, \tablefoottext{p}{\cite{Lillo-Box2014}}, \tablefoottext{q}{\cite{Wong2020b}}, \tablefoottext{r}{\cite{CollierCameron2010}}, \tablefoottext{s}{\cite{vonEssen2020}}, \tablefoottext{t}{\cite{Turner2016}}, \tablefoottext{u}{\cite{Sousa2018}}, \tablefoottext{v}{\cite{Delrez2016}}, \tablefoottext{w}{\cite{Bourrier2020}},\tablefoottext{x}{\cite{Lendl2020}}, \tablefoottext{y}{\cite{Deline2022}}, \tablefoottext{z}{\cite{Brandeker2022}}, \tablefoottext{$\alpha$}{\cite{Sousa2021}},  \tablefoottext{$\beta$}{\cite{Krenn2023}}}
\label{tab:albedos}
\end{table*}

\begin{figure}
   \centering
   \includegraphics[width=\hsize]{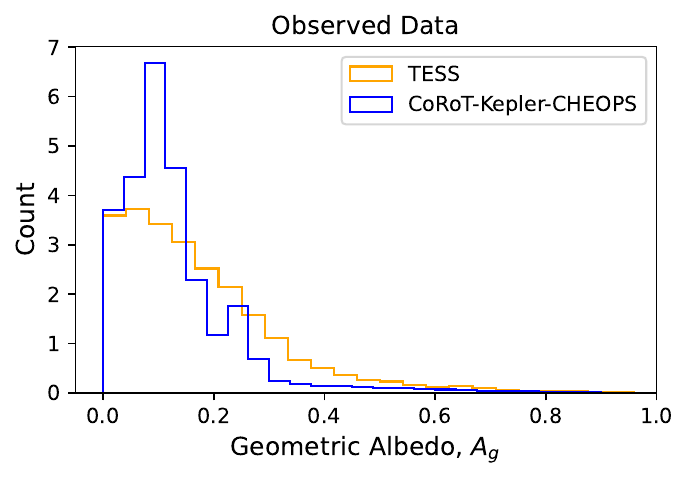}
      \caption{The distribution of observed geometric albedos for both bandpasses (CoRoT-Kepler-CHEOPS (CKC) in blue and TESS in orange), where each observation (from Table \ref{tab:albedos}) is treated as a Gaussian distribution with a width given by the 1-sigma uncertainty reported. This histogram plot is made from 10,000 samples of each observation and then normalised. The CKC sample peaks around $A_g = 0.1$ whereas the TESS sample has a wider peak and has a maximum at $A_g = 0$. However, this could be a result of the much wider errorbars in the TESS data, smoothing out any peaks.}
         \label{fig:ag_hist}
\end{figure}

The set of geometric albedos in each bandpass group is shown in Figure \ref{fig:ag_hist}. We draw 10,000 samples from the geometric albedo posterior distribution, assuming a Gaussian distribution with means and standard deviations defined by the observations and their uncertainty. Since the observations have large uncertainties that may extend to negative albedos, we take the absolute value before comparing to the model posteriors. The TESS albedo posterior samples peak at zero and the CKC samples peak around $A_g=0.1$. The CKC samples also have a narrower peak than the TESS samples, however this width will be very dependent on the uncertainties of the observations. This more positive peak of the bluer bandpasses is, however, in line with expectations, because scattering affects the blue end of the spectrum.  

\begin{figure*}
   \centering
   \includegraphics[width=\hsize]{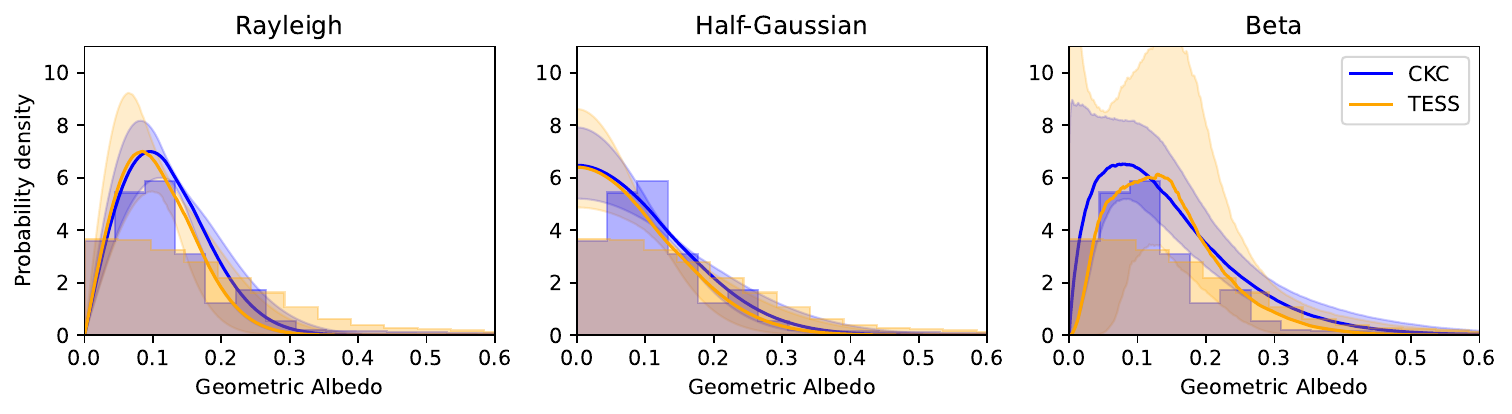}
      \caption{Three panels show the best-fit models (Rayleigh, half-Gaussian and Beta distributions) plotted over the observed geometric albedos, with the shading showing the $1\sigma$ uncertainties in the fits. In orange is the TESS band and in the blue, the CKC band. We find that the best-fit models from the two bandpasses are consistent. We also find that, by calculating the BIC of each fit, that the Rayleigh distribution is the best-fitting model for this data (although not very significantly) and the Beta distribution the worst. }
         \label{fig:distfit}
\end{figure*}

Following a similar approach as \cite{Sagear2023}, we compare these observed albedo distributions with three empirical albedo distribution models within a hierarchical Bayesian framework. The hypothetical, empirical albedo distributions are Rayleigh, half-Gaussian and Beta distributions, chosen for their simplicity, and to test whether the albedo distributions go to zero counts at zero albedo or not. The Rayleigh and half-Gaussian functions have only one free parameter, $\sigma$, and the Beta function has two parameters, $a$ and $b$. The likelihood functions for each distribution can be found in Appendix \ref{sec:likelihoods}. The best-fit models are shown in Figure \ref{fig:distfit} and Table \ref{tab:distbic} details the best-fit parameter values for the CKC and TESS dataset. 

\begin{table}[h]
\centering
\caption{Results of fitting different distributions to the observed geometric albedo data (see Figure \ref{fig:distfit}).}
\small
\begin{tabular}{lllc} 
 \hline
 \hline
 Distribution & Parameter & Best-fit value & BIC\\ 
 \hline
 Rayleigh & $\sigma_{\text{CH}}$ & $0.088\pm0.013$ & -336.2 \\
 & $\sigma_{\text{T}}$ & $0.089\pm0.021$ & -349.3 \\\\
 Half-Gaussian & $\sigma_{\text{CH}}$ & $0.128\pm0.027$ & -335.4 \\ 
  & $\sigma_{\text{T}}$ & $0.129\pm0.037$ & -348.7 \\\\
Beta & $(\log(a_{\text{CH}}),$ & $(0.264\pm0.209,$ & \\
  & $\;\log(b_{\text{CH}}))$ & $\;1.221\pm0.216)$ & -333.3\\\\
& $(\log(a_{\text{T}}),$ & $(0.509\pm0.528$ & \\
& $\;\log(b_{\text{T}}))$ & $\;1.488\pm0.496)$ & -346.3\\
\hline
\end{tabular}
\tablefoot{We fit independent distributions to the CKC and TESS datasets. Both Rayleigh and the half-Gaussian distribution have only 1 free parameter ($\sigma$) whereas the Beta distribution has two ($a, b$), which we fit for in log space to improve the efficiency of our sampling. The best-fit values show that all parameters are consistent across bandpass, indicating the distributions (CKC and TESS) are not significantly different. The BIC results show the three models fit the data similarly well, as the values are all within 10 points of each other. However we still do see, and one can confirm this by eye by looking at Figure \ref{fig:distfit}, that the Rayleigh distribution performs the best for both bandpass datasets, as it has the lowest BIC value. }
\label{tab:distbic}
\end{table}

We compare the Bayesian Information Criterion (BIC) between each albedo distribution model, and find that no model is strongly preferred. The Rayleigh function has the smallest BIC, for both CKC and TESS data, and the Beta function has the highest. Looking at the difference between the CKC and TESS distributions, we see that the best-fit parameters are consistent with each other. Again following \cite{Sagear2023}, we use the probability ratio:
\begin{equation}
    R = \frac{P(D_1D_2|H_1)}{P(D_1|H_0)P(D_2|H_0)},
\end{equation}
to evaluate whether a joint fit to the CKC and TESS datasets is preferred over individual fits. Here $D_1$ and $D_2$ represent the CKC and TESS geometric albedos and hypothesis $H_1$ states that a joint fit is best to fit these datasets, whilst $H_0$ states that individual fits to the datasets is best. $P(D_1|H_0)$ and $P(D_2|H_0)$ are the likelihoods of the best-fit models to the individual datasets and $P(D_1D_2|H_1)$ is the likelihood of the best-fit joint model. If a joint fit is preferred, this gives us evidence that the samples are from the same underlying distribution. For the Rayleigh model we obtain $\ln(R)=9.27$, for half-Gaussian $\ln(R) = 9.41$ and for Beta $\ln(R) = -19.1$. The two best-fitting models have positive values for $\ln(R)$, which implies that the joint fit likelihood is higher than the combined individual fits. We conclude there is no evidence for a difference in the underlying distributions of the TESS vs CKC-measured geometric albedos.

\begin{figure*}
   \centering
   \includegraphics[width=\hsize]{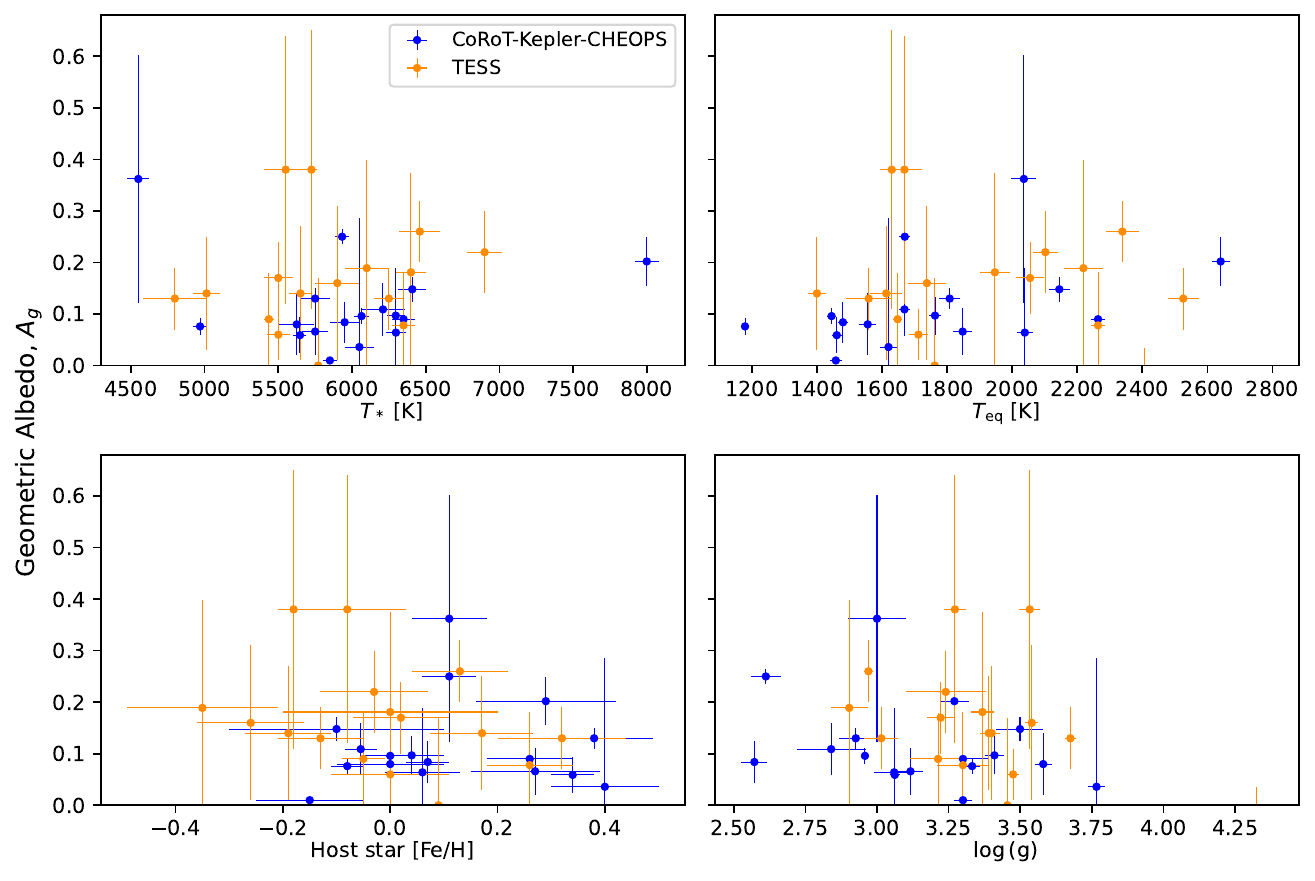}
      \caption{Geometric albedos and physical parameters for targets observed with CoRoT-Kepler-CHEOPS (blue points) and TESS (orange points). The trend results can be found in Table \ref{tab:bic}. We have restricted the y-axis to physical values of the geometric albedo, however it should be noted that some targets (WASP-18b and WASP-33b) have median posterior values below 0.} 
         \label{fig:allobs}
\end{figure*}

Next, we take the set of thermally decontaminated albedos and investigate whether they show any correlation with other system parameters. We show the albedos plotted against these parameters in Figure \ref{fig:allobs}. We fit both a constant and a linear model to the geometric albedos as a function of a range of parameters, taking into account the uncertainties in both the system parameters and the albedos, with \texttt{emcee} \citep{Hogg2010,Foreman-Mackey2013}. The albedo uncertainties are likely underestimated in this decontamination process, so we include an error bar scaling factor as a free parameter. These fits are intended for finding first-order trends -- we note that the linear fits allow non-physical, negative albedos. We show the BIC (Bayesian Information Criterion) results of these model comparisons in Table \ref{tab:bic}.

\begin{table*}[h]
\centering
\caption{Result of the trend analysis of the observed geometric albedos versus various stellar/planetary parameters ('Independent Variable' in this table).}
\begin{tabular}{ c|lccc } 
 \hline
 \hline
 Independent Variable & Band & Linear model BIC & Constant model BIC & $\Delta$ BIC \\ 
 \hline
 $T_\star$ & TESS & -49.0 & -48.4 & 0.6 \\ 
 & CKC & -38.0 & -40.2 & -2.2 \\
 \hline
 $T_{\textrm{eq}}$ & TESS & -48.9 & -48.4 & 0.50\\ 
 & CKC & -41.8 & -40.2 & 1.6 \\
 \hline
Log(g) & TESS & -54.4 & -48.4 & 6.0 \\
& CKC & -52.3 & -40.2 & 12.1 \\
\hline
[Fe/H] & TESS & -46.7 & -48.5 & -1.8 \\
& CKC & -49.3 & -40.2 & 9.1
\end{tabular}
\tablefoot{As explained in Section \ref{sec:obs}, we fit two different trends to the geometric albedo data: a constant and a linear trend, considering uncertainties in both the albedo and free parameter. We calculated the BIC difference between these two models for each free parameters and for each bandpass dataset. Typically, a $\Delta {\rm BIC} < 10$ is considered not significant. In general, we find no significant trends. There is a slight preference for a linear model across the different variables, however since the $\Delta BIC$ is mostly below 10, this is not significant. If the model can't choose between a constant or a linear trend, then we conclude there is little evidence to suggest the data shows a correlation. The only combination where there is a $\Delta BIC > 10$ is the CKC albedos as a function of $\log(g)$. Physically, this could be interesting, however, looking at the data, it is clear that the trend is mostly driven by the few outer points at the low and high end of the log(g) space. At this point, more precise data across the full range of the free parameter is needed to confirm this correlation.}
\label{tab:bic}
\end{table*}

The uncertainties of the geometric albedos are large compared to the range of possible values (between 0 and 1). In general, the uncertainties for the TESS measurements have larger error bars than the CKC measurements. This is partially due to the error on the reported eclipse depth. Most of the Kepler objects have very tight constraints on their eclipse depths, due to the high number of eclipses per object. The slopes of these linear fits are sensitive to the measurements at the extremes. Typically $\Delta$BIC$<$10 between constant and linear models, suggesting no preference for correlations. There is a weak preference for the linear model with $\log(g)$ and [Fe/H] in the CKC band.

Whilst the albedo measurements show little or no correlations with system parameters, the geometric albedo measurements are very uncertain. Another reason for this uncertainty comes from the thermal decontamination procedure, which often involves multiple assumptions (e.g. blackbody planets) and very uncertain dayside temperature measurements due to limited infrared observations. Emitted flux scales with $\textrm{T}^4$ (within a specific bandpass this relation is not as straightforward), so these uncertainties propagate through to the geometric albedo. A more correct thermal decontamination approach would use the full planetary emission spectrum, but almost none of these planets, to date, have full SED measurements from, e.g., JWST.

\section{Theoretical model results}
\label{sec:model}
\subsection{Geometric albedos in chemical equilibrium}
\label{sec:chemeq}

We began our modelling efforts by assuming chemical equilibrium in the planet's atmosphere (see Section \ref{sec:modellingAg} for model details). Table \ref{tab:chemeq_priors} details the priors used for the sampling distributions. We sampled over $T_\star$, $T_{\textrm{planet}}$, pressure and metallicity (i.e. the observed atmospheric metallicity). 

\begin{table}[h]
\centering
\caption{Priors distributions for the input to our geometric albedo model assuming chemical equilibrium. This model includes $\text{H}_2\text{O}$ and Na.}
\begin{tabular}{ |c|c| } 
 \hline
 Parameter & Assumed distribution \\ 
 \hline
 $T_\star$ & $\mathcal{U}(4550, 8000)$\,K \\ 
 $T_{\textrm{planet}}$ &  $\mathcal{U}(1300, 2700)$\,K\\ 
 pressure & $\log\mathcal{U}(-3, -1)\,\textrm{bar}$ \\
 $[M/H]$ & $\log\mathcal{U}(-1,1)$ \\
 \hline
\end{tabular}
\label{tab:chemeq_priors}
\end{table}

\begin{figure}
   \centering
   \includegraphics[width=0.45\textwidth]{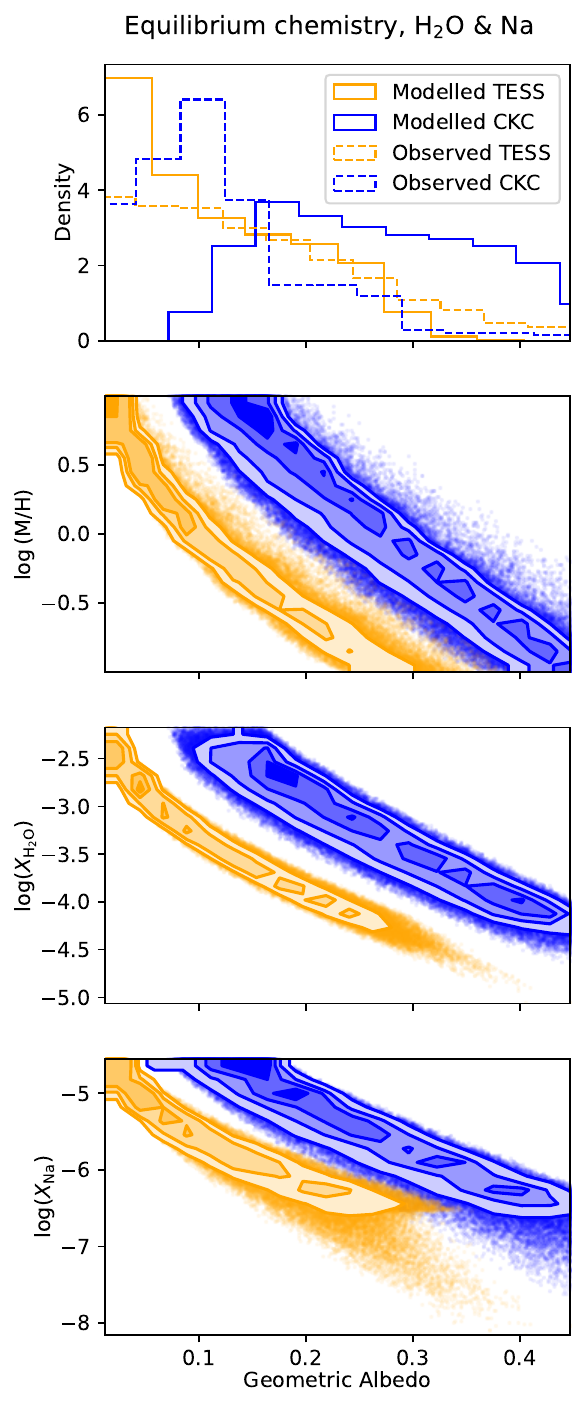}
      \caption{Geometric albedo model results, produced by sampling over the input parameter prior distributions (see Table \ref{tab:chemeq_priors}) and assuming chemical equilibrium. The wavelength-dependent albedos were then bandpass-integrated to produce the expected albedos in the TESS (orange) and CKC (blue) band. Not shown are the albedos as a function of $T_*$, $T_{\text{planet}}$ and pressure, as these show no discernible correlation. We find that the metallicity (log [M/H]) has the largest impact on the geometric albedo and that from this, water abundance produces slightly tighter constraints on $A_g$ than sodium. We also see very distinct differences between the CHEOPS and TESS bandpasses.}
         \label{fig:corner_chemeq}
\end{figure}

Figure \ref{fig:corner_chemeq} shows the results in the CHEOPS and TESS bandpass. Abundances of the species are not free parameters themselves, but are calculated from the M/H ratio, pressure and temperature. As we have already mentioned, the use of a single number (the metallicity) to describe the gas-phase chemical abundances is a simple assumption. The metallicity is correlated with the geometric albedo due to larger abundances of absorbing molecules increasing the opacity in the atmosphere. From our results it is clear that water and sodium have strong and similar effects on suppressing the albedo. For the same chemical abundances, we generally measure a lower $A_g$ in the TESS bandpass than the CHEOPS bandpass. Interestingly, the albedo distributions from the different bandpasses have little overlap. The most reflective planet in the TESS bandpass almost always has a lower albedo than the least reflective planet in CHEOPS bandpass. This is not borne out by the observations. We do not show the albedos as a function of $T_*$, $T_{\text{planet}}$ and pressure, as these show no discernible correlation.

\subsection{Geometric albedos out of equilibrium}

\begin{figure}
   \centering
   \includegraphics[width=0.45\textwidth]{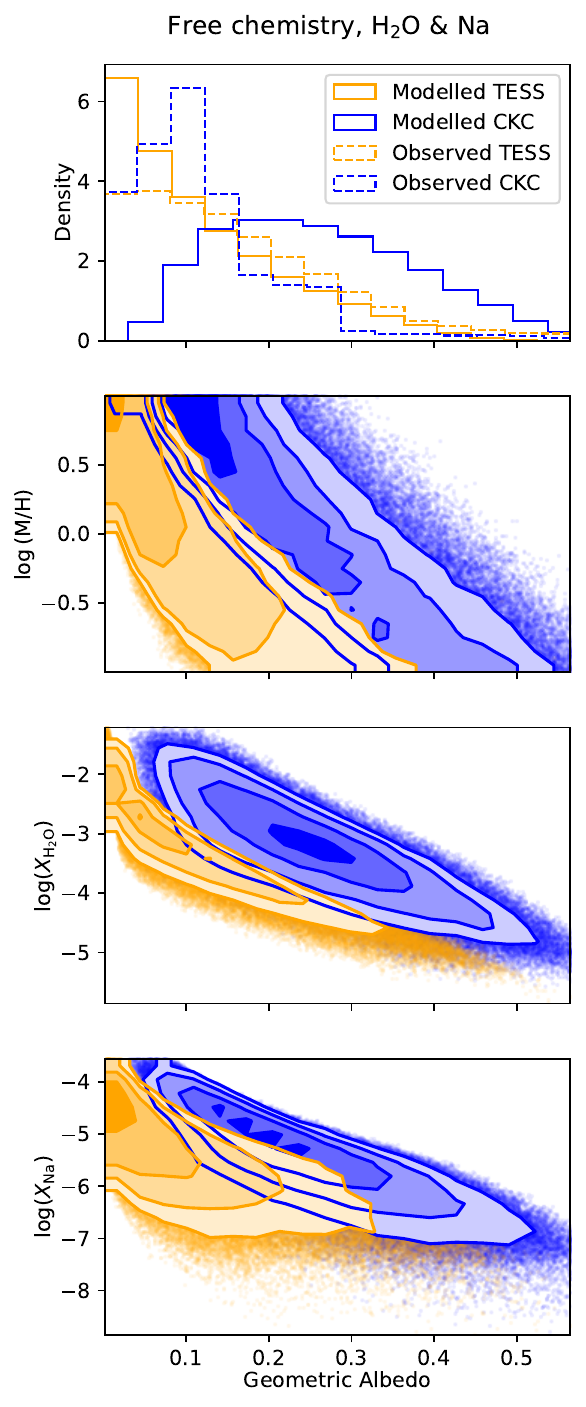}
      \caption{Geometric albedo model results, produced by sampling over the input parameter prior distributions (see Table \ref{tab:nochemeq_priors}) and using a scaling factor distribution to move the model away from chemical equilibrium. The wavelength-dependent albedos were then bandpass-integrated to produce the expected albedos in the TESS (orange) and CKC (blue) band. We find similar results to the chemical equilibrium model: temperature and pressure had no effect on the geometric albedo (so they are not shown here), and metallicity is the main driver. Due to increased parameter space from the relaxation of the chemical equilibrium constraint, the posteriors of TESS and CKC albedos are now wider and begin to overlap slightly. It is clear that the higher the metallicity, the lower the geometric albedo.   }
         \label{fig:corner_nonchemeq}
\end{figure}

Next, we relax the assumption that the planet's atmosphere is in chemical equilibrium, using the priors in Table \ref{tab:nochemeq_priors}. To parameterise the disequilibrium, we use a scaling factor to perturb the mixing ratios of each species relative to the equilibrium value. We independently sample over a distribution of 'scaling factors' which has the form of $\log\mathcal{U}[-1,1]$ in log space for both sodium and water. Our results are shown in Figure \ref{fig:corner_nonchemeq}. We find similar trends as in Section \ref{sec:chemeq}, however with a larger scatter of $A_g$ for a fixed chemical abundance.

\begin{table}[h]
\centering
\caption{Priors distributions for the input to our geometric albedo model with chemical abundances scaling factors to shift the model away from chemical equilibrium. This model includes $\text{H}_2\text{O}$ and Na.}
\begin{tabular}{ |c|c| } 
 \hline
 Parameter & Assumed distribution \\ 
 \hline
 $T_\star$ & $\mathcal{U}(4550, 8000)$\,K \\ 
 $T_{\textrm{planet}}$ &  $\mathcal{U}(1300, 2700)$\,K\\ 
 pressure & $\log\mathcal{U}(-3, -1)\,\textrm{bar}$ \\
 $[M/H]$ & $\log\mathcal{U}(-1, 1)$ \\
 scale $X_{\text{H}_2\text{O}}$ & $\log\mathcal{U}(-1, 1)$ \\
 scale $X_{\text{Na}}$ & $\log\mathcal{U}(-1, 1)$ \\
 \hline
\end{tabular}
\label{tab:nochemeq_priors}
\end{table}

Due to the increased scatter, the values of $A_g$ in the TESS and CHEOPS bandpass are permitted to overlap. However it is still clear that, on average, the albedos from CHEOPS are higher than from TESS. From the width of the posterior distributions, we observe that water has a stronger influence than sodium in the TESS bandpass and they both have similar-strength correlations in the CHEOPS band. 

\subsection{Does the theoretical model match the observations?}

\begin{figure}
   \centering
   \includegraphics[width=\hsize]{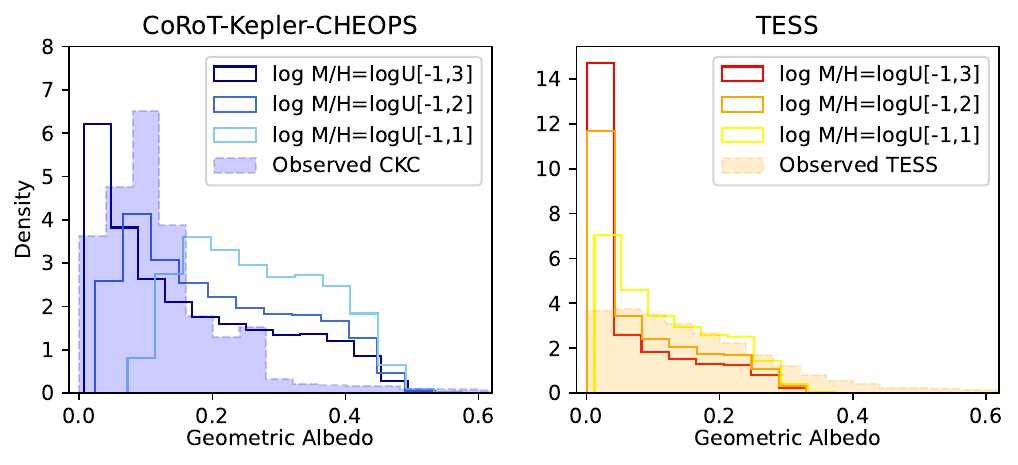}
      \caption{The assumed priors for the underlying abundances of water and sodium have a large influence on the modelled geometric albedo. On the left panel we show the distributions in the CHEOPS bandpass, and on the right we show the distributions in the TESS bandpass. We overlay the observed distribution to suggest perhaps the underlying effective metallicity distribution is within these ranges, however we are also very aware that the uncertainties of the observations highly influence this shape (due to the low number of observations). }
         \label{fig:comparingmodelwithdatametallicity}
\end{figure}

\begin{figure}
   \centering
   \includegraphics[width=\hsize]{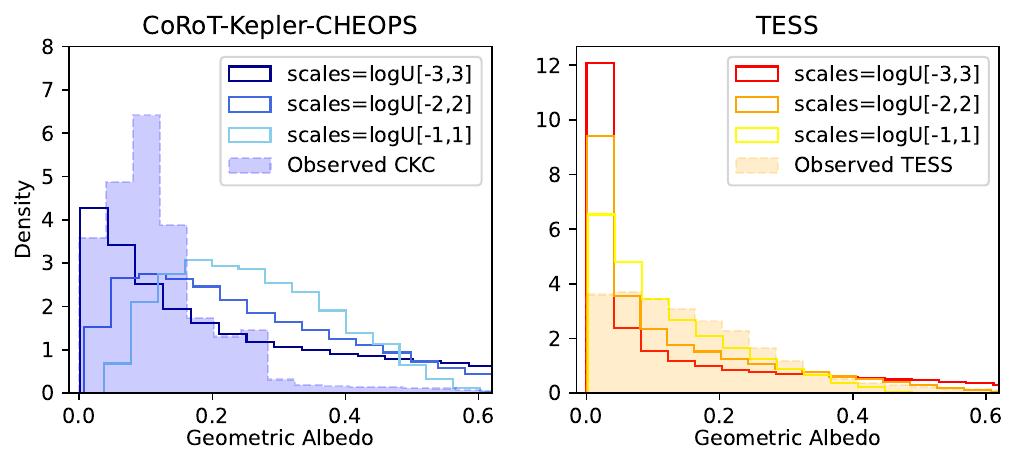}
      \caption{How far away from equilibrium the abundances of water and sodium are permitted to go have a large influence on the modelled geometric albedo. This is mostly due to the fact larger metallicities can be explored with a larger scaling prior. On the left panel we show the distributions in the CHEOPS bandpass, and on the right we show the distributions in the TESS bandpass. Here we keep the metallicity distribution fixed to $\log\mathcal{U}[-1,1]$. We overlay the observed distribution. }
         \label{fig:comparingmodelwithdatascaling}
\end{figure}

We reproduce the range of albedos observed with TESS and CKC using a model with only $\text{H}_2$, water and sodium. In our model simulations, the very low metallicity planets produce high albedos, in both the CHEOPS and TESS band, because there are too few absorbers to stop the Rayleigh scattering from dominating. This strong  correlation between albedo and metallicity is not mirrored in the observations, if we assume the stellar metallicity is equal or correlated to the planet's metallicity. However there are huge uncertainties on the stellar metallicity measurements, making it very difficult to detect any trends. From planet formation, stellar metallicity and planetary metallicity share no simple linear relation, especially if there is condensation involved which acts to remove metals from the gas phase. We also find no significant correlation with the stellar temperature, planetary temperature or pressure. Over the range considered, these parameters only have a very small effect on the cross-section of the chemical species and so it is expected that they should not have a significant influence on $A_g$.

The comparison between the theoretical and observed distributions is very informative. We find both similarities and differences between the shapes of the distributions, for example, the peak of the modelled CKC results is at a higher albedo than the TESS distribution. Although this matches what we find in the data, we believe the observed sample size is too small to trust this as a result. More observations would be needed to confirm this hypothesis. We also find that, due to the metallicity constraints, the CKC modelled distribution does not reach zero. In Figure \ref{fig:comparingmodelwithdatametallicity}, we show how the distributions depend on the assumed metallicity prior. The higher the metallicities probed, the greater a pile-up of albedos at zero. In Figure \ref{fig:comparingmodelwithdatascaling}, we show how increasing the distance the model can probe away from chemical equilibrium changes the shape of the modelled distributions. This is mostly due to the fact that larger metallicities can be explored with a larger scaling prior.

\subsection{Theoretical models with more absorbers}
Based on the analogy to brown dwarfs, \cite{Fortney2008} predicted the existence of two classes of hot Jupiters based on the absence or presence of TiO and VO. Since then, there has been a handful of confirmed detections of TiO or VO in hot Jupiters to date \citep[see, e.g.][]{Nugroho2017,Chen2021}. It is known that TiO and VO are prominent absorbers in hot atmospheres when observed in emission and are also detectable at high resolution via cross-correlation. Additionally, TiO and VO have been the proposed absorbers responsible for detected thermal inversions \citep[see, e.g.][]{Coulombe2023, Cont2021}. However, these species have not been detected directly in reflection spectroscopy, which would be the only way to conclusively determine the albedos are affected directly by TiO/VO in an atmosphere. One possible reason is the sequestration of titanium and vanadium into condensates by a ``cold trap" on the nightside of hot Jupiters \citep[see, e.g.][]{Parmentier2013}. If they are condensed, then they would not contribute to the stellar absorption and scattering in the atmosphere. 

\begin{figure}[h]
   \centering
   \includegraphics[width=\hsize*8/10]{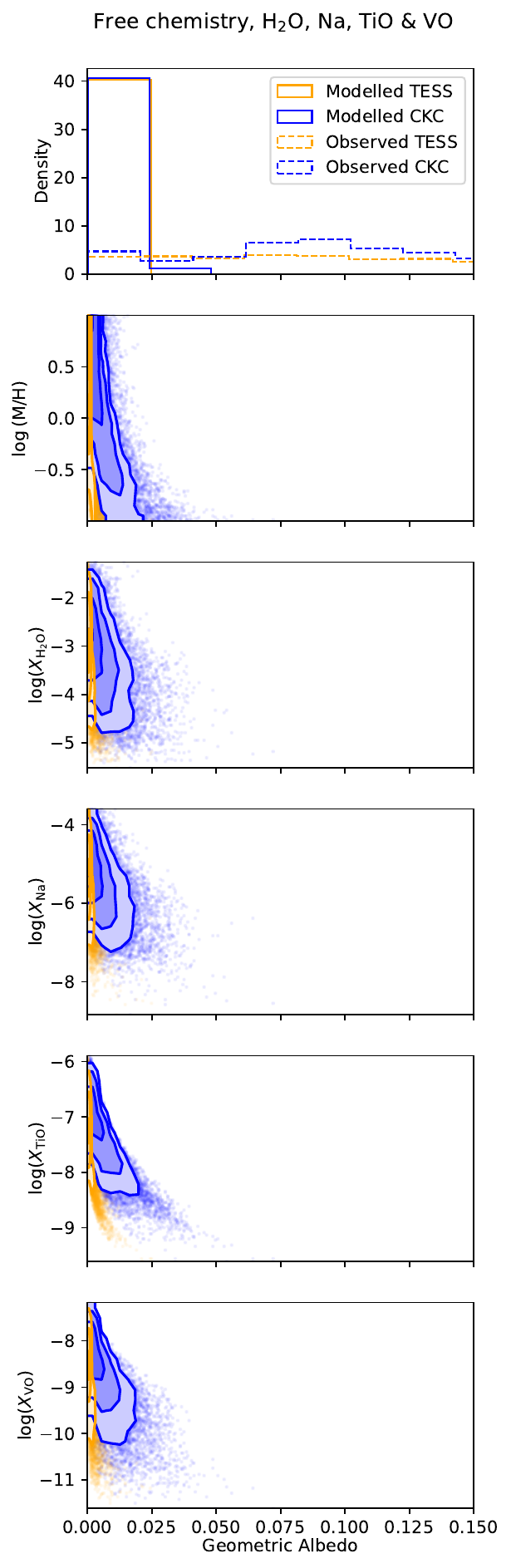}
      \caption{Geometric albedo model results without condensation, produced by sampling over the input parameter prior distributions (see Table \ref{tab:morespecies_priors}) and using a scaling factor distribution to move the model away from chemical equilibrium. The wavelength-dependent albedos were then bandpass-integrated to produce the expected albedos in the TESS (orange) and CKC (blue) band. We find that the presence of additional optical absorbers forces the albedo to stay near zero, even when the metallicity is low. }
         \label{fig:corner_morespecies}
\end{figure}

\begin{figure}[h]
   \centering
   \includegraphics[width=\hsize*7/9]{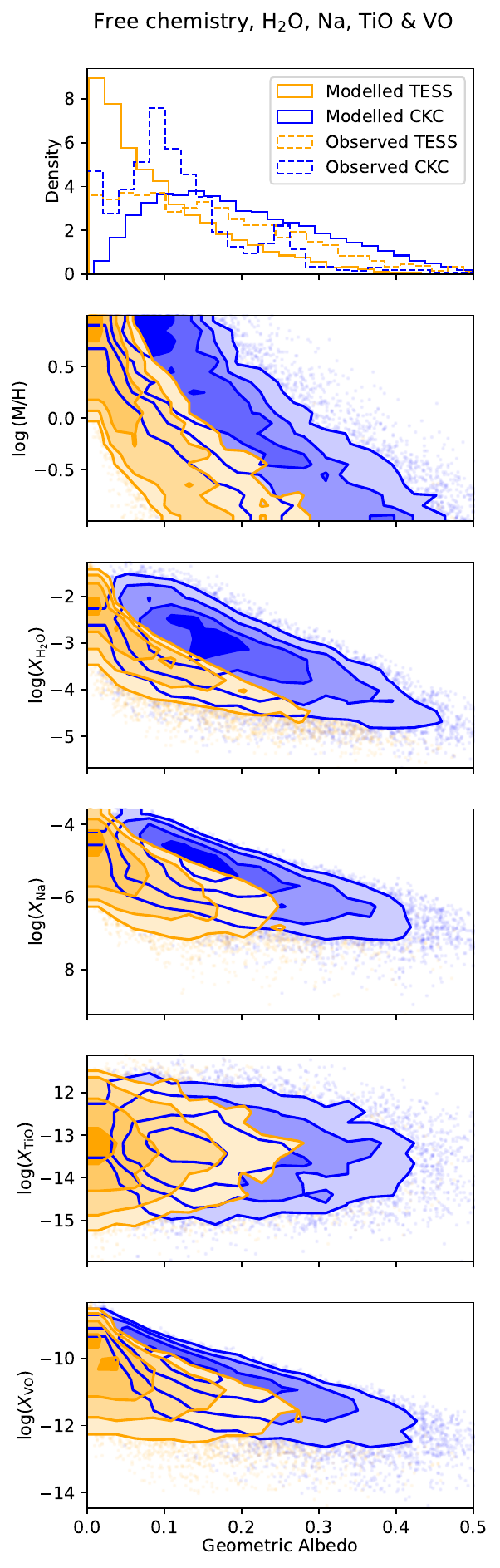}
      \caption{Geometric albedo model results including condensation, produced by sampling over the input parameter prior distributions (see Table \ref{tab:morespecies_priors}) and using a scaling factor distribution to move the model away from chemical equilibrium. The wavelength-dependent albedos were then bandpass-integrated to produce the expected albedos in the TESS (orange) and CKC (blue) band. We find that high albedos are reachable with high metallicites, due to condensation removing the effect of metals from the upper atmosphere.}
         \label{fig:corner_morespecies_withcondensation}
\end{figure}

This is to say, that even if the initial abundances of TiO and VO are correctly predicted by gas-phase equilibrium chemistry, their final abundances are controlled by atmospheric dynamics and condensation. Therefore, we know that these species could be present, but we do not know exactly where or in what abundance, and consequently how much they contribute to the geometric albedo we observe. This motivated us to test atmospheres with and without these species in this study, using a scaling factor relative to chemical equilibrium. We also tested adding condensation into our model to see what effect this would have. When we use the model without condensation, we do not let the scaling factors of TiO or VO be greater than 1 (or 0 in log-space), to maintain physical conditions. For the other species and parameters, we use similar priors to the other models (see Tables \ref{tab:morespecies_priors} and \ref{tab:morespecies_priors_withcondensation}). Our results are shown in Figures \ref{fig:corner_morespecies} and \ref{fig:corner_morespecies_withcondensation}. We see that in the absence of a condensation model, including these strong-optical absorbers, even at small abundances, drastically reduces the observed geometric albedo. However, when the condensation is switched on, we find higher geometric albedos are again reachable. We find that these distributions are very different from the observed distribution, in both the CHEOPS and TESS bandpass.

\begin{table}[h]
\centering
\caption{Priors distributions for the input to our geometric albedo model with chemical abundance scaling factors to shift the model away from chemical equilibrium. This model includes $\text{H}_2\text{O}$, Na, TiO and VO. }
\begin{tabular}{ |c|c| } 
 \hline
 Parameter & Assumed distribution \\ 
 \hline
 $T_\star$ & $\mathcal{U}(4550, 8000)$\,K \\ 
 $T_{\textrm{planet}}$ &  $\mathcal{U}(1300, 2700)$\,K\\ 
 pressure & $\log\mathcal{U}(-3, -1)\,\textrm{bar}$ \\
 $[M/H]$ & $\log\mathcal{U}(-1, 1)$ \\
 scale $X_{\text{H}_2\text{O}}$ & $\log\mathcal{U}(-1, 1)$ \\
 scale $X_{\text{Na}}$ & $\log\mathcal{U}(-1, 1)$ \\
 scale $X_{\text{TiO}}$ & $\log\mathcal{U}(-1, 0)$ \\
 scale $X_{\text{VO}}$ & $\log\mathcal{U}(-1, 0)$ \\
 \hline
\end{tabular}
\label{tab:morespecies_priors}
\end{table}

\begin{table}[h]
\centering
\caption{Priors distributions for the input to our geometric albedo model with condensation enabled and chemical abundance scaling factors to shift the model away from chemical equilibrium. This model includes $\text{H}_2\text{O}$, Na, TiO and VO. }
\begin{tabular}{ |c|c| } 
 \hline
 Parameter & Assumed distribution \\ 
 \hline
 $T_\star$ & $\mathcal{U}(4550, 8000)$\,K \\ 
 $T_{\textrm{planet}}$ &  $\mathcal{U}(1300, 2700)$\,K\\ 
 pressure & $\log\mathcal{U}(-3, -1)\,\textrm{bar}$ \\
 $[M/H]$ & $\log\mathcal{U}(-1, 1)$ \\
 scale $X_{\text{H}_2\text{O}}$ & $\log\mathcal{U}(-1, 1)$ \\
 scale $X_{\text{Na}}$ & $\log\mathcal{U}(-1, 1)$ \\
 scale $X_{\text{TiO}}$ & $\log\mathcal{U}(-1, 1)$ \\
 scale $X_{\text{VO}}$ & $\log\mathcal{U}(-1, 1)$ \\
 \hline
\end{tabular}
\label{tab:morespecies_priors_withcondensation}
\end{table}

\section{Discussion}

\subsection{Geometric albedos are primarily determined by the abundance of absorbers}

Using our geometric albedo model (Section \ref{sec:modellingAg}), we have shown evidence that the abundances of absorbing chemical species in the atmosphere (or 'effective' metallicity) is one of the main drivers for determining the geometric albedo of a planet. We have also seen, that when heavier metals such as TiO and VO are present, the efficiency of condensation alongside the abundances are the main drivers. From Figure \ref{fig:corner_morespecies}, we see that when (optical) absorbers are abundant in the atmosphere, the (optical) geometric albedo is low. The exact correlation depends on which absorbers are present and whether the atmosphere is near chemical equilibrium or not.

The other potential correlates that we considered were atmospheric temperature, pressure and stellar temperature. We found no significant correlation between these parameters and the geometric albedo, within the ranges set by our observations. This is because the opacities of the atmospheric species do not vary significantly within this temperature and pressure range. Temperature has more of an impact on the volume-mixing ratio, however, but the change is insufficient to affect the bandpass-integrated albedo. One caveat here is that the model does not assume a correlation between stellar temperature and planet temperature, as this is beyond the scope of the model. Another caveat is that our model does not feature clouds, which, if present, may be more dependent on temperature and pressure and therefore influence the geometric albedo more significantly.

\subsection{What this means for atmospheric characterisation}
\label{sec:flowchart}
We have summarised our model results into a simple framework for interpreting albedo measurements in terms of atmospheric properties. In broad terms:

\begin{itemize}
    \item First, measure the optical geometric albedo of the planet.
    \item If the albedo is high ($A_g \gtrsim0.2$), then the planetary atmosphere likely produces significant scattering (to scatter the starlight back into space), or it has few absorbers (allowing Rayleigh scattering to dominate).
    \item If the albedo is low ($A_g \lesssim0.05$), then this could imply the atmosphere has little scattering, or it is abundant in absorbers. 
    \item If the albedo is high and the metallicity is high, significant additional scattering in the atmosphere is required to produce the observed albedo. None of our models (with different absorbers or abundance combinations) have been able to produce high geometric albedos when the metallicity is high. Reflective clouds would be an obvious solution to produce the additional scattering required. 
    \item High albedos can be produced by planets with low metallicity if they have only only water and sodium present. Adding more absorbers such as TiO and VO causes the albedo to decline rapidly without the addition of clouds. This scenario is sensitive to the absorbers in the atmosphere.  
    \item Alternatively, if the albedo is low and the metallicity is high, abundant absorbers must dominate over the Rayleigh scattering in the atmosphere. \textbf{This scenario may also imply that clouds are not present in this atmosphere}, since clouds tend to boost the geometric albedo.
    \item The final combination is a low geometric albedo and low metallicity, which, we have shown in our water and sodium models, is impossible unless the chemistry strays far from equilibrium.
\end{itemize}

\subsection{Which parameters control the spread in $A_g$?}
Figure \ref{fig:corner_chemeq} shows that the model albedo distributions with equilibrium chemistry have higher probability density near $A_g = 0$ than the CKC observations, which peak $A_g = 0.2$. The observed TESS albedos may lack this peak because the albedo uncertainties are much larger.

The model geometric albedo distributions increase towards $A_g\rightarrow0$ mostly due to high metallicities. Perhaps the highest metallicities in this prior are excluded by the observations, and may not occur in nature. Looking at the stellar metallicities associated with the observations, they are all below [Fe/H] $\sim 0.4$. At present, devising any further parametrisation to relate stellar metallicity and planet metallicity might be premature.

We note that the spread in geometric albedos is larger with disequilibrium chemistry. In equilibrium, the geometric albedos observed in the two bandpasses do not overlap. This is not something that is seen in the observations.

\subsection{Albedo differences between bandpasses}

For the model in chemical equilibrium with only water and sodium, we see large discrepancies between the TESS and the CHEOPS bandpass, despite significant wavelength overlap between the bandpasses. Even with our generous priors allowing mixing ratios to range from 0.1-10 times the equilibrium abundance (and more, see Figure \ref{fig:comparingmodelwithdatascaling}), the posteriors are clearly still distinct. 

However, we found that observed albedos from TESS and CKC are very similar. In Section \ref{sec:obs}, we curated geometric albedos in the literature and concluded the CKC and TESS observed albedo datasets come from the same underlying distribution. There are two likely reasons why the observed albedo distributions are mutually consistent, but the model albedo distributions are inconsistent. The first is that the observational uncertainties are too large to identify small differences. The second is that the model is quite simplified. At planetary temperatures near 2000\,K, we expect condensation of clouds to play an important role in the observed albedos, which may obscure any differences between the albedo distributions from CKC and TESS. 

\subsection{Future work}

With the help of CHEOPS and Kepler, we have very precise eclipse depth measurements, giving us the total planetary flux in a specific bandpass. The problem is how to decouple the reflected light from the thermal emission of the planet. From measurements of the solar spectrum \citep[e.g.][]{Fraunhofer1918}, we know that approximating stars and planets as black bodies is a bad approximation. The step-up from this is to fit a planetary model spectrum to both Spitzer infrared eclipse depths and then extrapolate this model into the optical. Of course, two data points is not ideal for fitting a full atmospheric spectrum. The most 'correct' method would be to take a full emission spectrum of the planet and jointly fit a reflected light component and a thermal emission component to obtain the true $A_g$. With the James Webb Space Telescope (JWST) optical-to-IR modes such as NIRSpec prism, we can begin to effectively decontaminate secondary eclipses. The Nancy Grace Roman Space Telescope is set for launch in 2027, which may increase the number of know planets by over an order of magnitude \citep{Wilson2023, Tamburo2023}, and enable direct imaging in reflected light \citep[see, e.g.][]{Carrion-Gonzalez2021}, giving us a more direct method to separate the thermal and reflected light. 

Furthermore, the community has been working for over a decade to understand cloud formation in these planets, and this work will be crucial for creating more accurate atmospheric models. As we see on Earth, clouds reflect significant amounts of solar light and so it is reasonable to assume they would have an impact on the scattering in a hot Jupiter's atmosphere. In Section \ref{sec:flowchart}, we discussed cases where clouds were and were not necessary to explain the derived geometric albedo, given a certain metallicity. Future work should now include trying to integrate a cloud model within a normal atmospheric model to estimate geometric albedos of planets. However this also depends on knowing the chemical compositions of a planet's atmosphere, something that is now becoming possible for these larger planets with JWST.

\section{Conclusion}

In this paper, we have conducted an investigation into the geometric albedos of hot Jupiters, comparing observational data from several space telescopes—TESS, Kepler, CoRoT, and CHEOPS—with theoretical models. Our findings demonstrate that the geometric albedo of these exoplanets is primarily influenced by the abundance of atmospheric absorbers, with metallicity playing a crucial role. The observed albedo distributions across different bandpasses do not exhibit significant differences. This differs from the models, where, if chemical equilibrium is assumed, there is a clear separation between the geometric albedos in the TESS bandpass compared to the CHEOPS bandpass. However, there are large uncertainties on the observed datapoints which could be masking real structure in the distributions which is predicted by the models. 

Our modelling efforts highlight the challenges in accurately predicting geometric albedos, especially when deviating from chemical equilibrium. The inclusion of additional absorbers like TiO and VO in our models consistently results in lower albedos, emphasising the complexity of these atmospheres and how much the specific absorbers influence the reflectivity of the planet.

Looking forward, the introduction of more precise instruments, particularly JWST and the Nancy Grace Roman Space Telescope, will be influential in refining our understanding of these albedo measurements by providing more accurate spectral data. This will allow for better separation of reflected and thermal emission components and precision atmospheric composition measurements of hot Jupiter atmospheres. Our work lays the groundwork for future studies to further explore the connection between atmospheric composition and reflectivity in exoplanet atmospheres.

\bibliographystyle{aa} 
\bibliography{bibliography} 

\appendix

\section{Likelihood distributions}
\label{sec:likelihoods}
Here we detail the likelihood distributions used to fit the Rayleigh, half-Gaussian and Beta distribution to the distribution of the observed geometric albedo data. $K$ is the total number of $A_g$ observations and $N$ is the number of samples of each observation distribution. 

\begin{equation}
\log\mathcal{L}_{\text{Rayleigh}} =  -\log(N) + \sum_{k=1}^{K} \log\left[ \sum_{n=1}^{N} \frac{{A_g}_k^n}{\sigma^2} \exp{\left( \frac{-{{A_g}_k^n}^2}{2\sigma^2}\right)}\right]
\end{equation}
\begin{equation}
\log\mathcal{L}_{\text{half-Gauss}} =  -\log(N) + \sum_{k=1}^{K} \log\left[ \sum_{n=1}^{N} \exp{\left( \frac{-{{A_g}_k^n}}{2\sigma}\right)^2} \frac{\sqrt{2}}{\sigma\sqrt{\pi}}\right]
\end{equation}
\begin{equation}
\log\mathcal{L}_{\text{Beta}} =  -\log(N) + \sum_{k=1}^{K} \log\left[ \sum_{n=1}^{N} \frac{\Gamma(a+b)({A_g}_k^n)^{a-1} (1-{A_g}_k^n)^{b-1}}{\Gamma(a)\Gamma(b)}\right]
\end{equation}

\section{Testing the effect of additional absorption in \textit{Spitzer}}
\label{sec:newalbedos}
\begin{figure}
   \centering
   \includegraphics[width=\hsize]{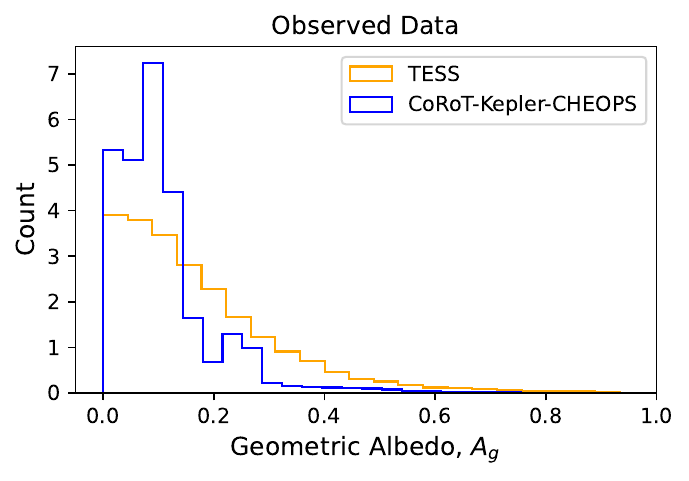}
      \caption{The distribution of observed geometric albedos for both bandpasses (CoRoT-Kepler-CHEOPS (CKC) in blue and TESS in orange), similar to Figure \ref{fig:ag_hist}, but with some results modified to account for potential absorption in the Spitzer band (as explained in Section \ref{sec:newalbedos}. This histogram plot is made from 10,000 samples of each observation and then normalised. }
         \label{fig:newalbedosdist}
\end{figure}

\begin{figure*}
   \centering
   \includegraphics[width=\hsize]{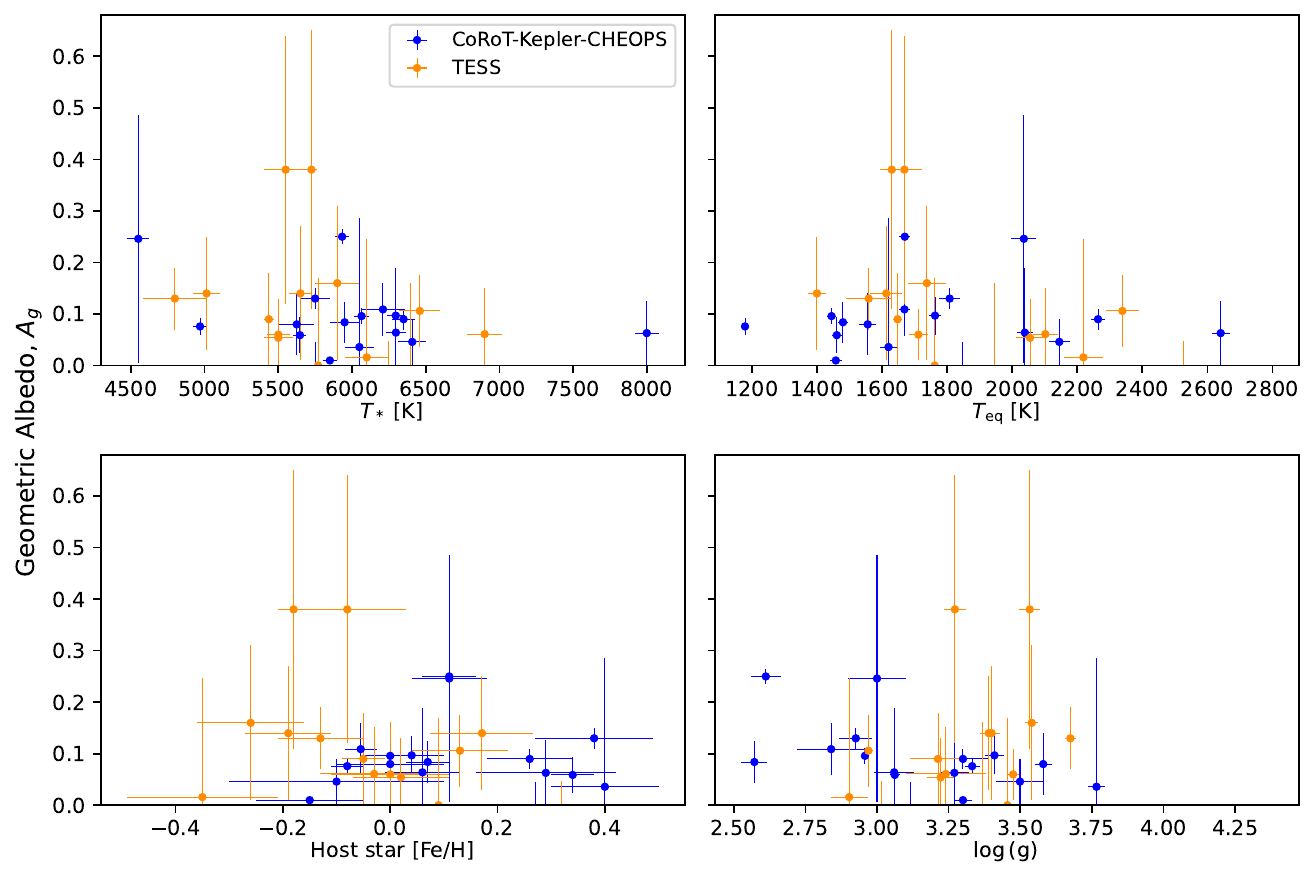}
      \caption{Geometric albedos and physical parameters for targets observed with CoRoT-Kepler-CHEOPS (blue points) and TESS (orange points), but with some results modified to account for potential absorption in the Spitzer band (as explained in Section \ref{sec:newalbedos}. We have restricted the y-axis to physical values of the geometric albedo, however it should be noted that some targets have median posterior values below 0.}
         \label{fig:newalbedosplots}
\end{figure*}

Brightness temperatures vary across bandpasses, so using the thermal decontamination approach from \textit{Spitzer} to TESS may neglect real differences in temperature with altitude. We tested the sensitivity of the model albedos with respect to changes in the near-infrared brightness temperature, as in \cite{Piette2020}. We added 250\,K to the Spitzer brightness temperature for all planets with $T_{eq} > 1800\,K$ to investigate the impact of the Spitzer decontamination approach on the albedos. In Figure \ref{fig:newalbedosdist}, we show a version of Figure \ref{fig:ag_hist} that has the added temperature offset, and in Figure \ref{fig:newalbedosplots}, we show Figure \ref{fig:allobs} with the same offset. 

\end{document}